# TRANSVERSE INSTABILITY OF THE ANTIPROTON BEAM IN THE RECYCLER RING*

L.R. Prost#, C.M. Bhat, A. Burov, J. Crisp, N. Eddy, M. Hu†, A. Shemyakin,
FNAL, Batavia, IL 60510, USA.

*Abstract*

The brightness of the antiproton beam in Fermilab's 8 GeV Recycler ring is limited by a transverse instability. This instability has occurred during the extraction process to the Tevatron for large stacks of antiprotons even with dampers in operation. This paper describes observed features of the instability, introduces the threshold phase density to characterize the beam stability, and finds the results to be in agreement with a resistive wall instability model. Effective exclusion of the longitudinal tails from Landau damping by decreasing the depth of the RF potential well is observed to lower the threshold density by up to a factor of two.

## INTRODUCTION

The Recycler ring (RR) is the last (third) ring in a chain of antiproton cooling and stacking stages. Transverse instabilities in RR have been theoretically studied during its design but were deemed a marginal issue for the maximum number of antiprotons that were expected to be stored at any time ($< 250 \times 10^{10}$ [1]). With strong electron cooling and up to $5 \times 10^{12}$ stored antiprotons, much brighter beams than initially anticipated are generated. As a result, emittances of the cooled beam are limited by a transverse instability [2]. A damper system was installed in 2005 with an initial bandwidth of 30 MHz [3], and eventually upgraded to 70 MHz [4]. Nevertheless, several instabilities were observed during normal operation and prompted studies to better understand their nature and characteristics, as well as to limit their occurrences.

## INSTABILITY MODEL

*Theory overview*

Since the antiprotons are accumulated within long bunches with low synchrotron frequency, a rigid beam model [5] is a reasonable approximation to the stability problem. Let us consider a coasting beam with Gaussian distributions in all planes and assume the chromaticity to be the main reason for the frequency spread, and introduce the effective phase density as,

$$D_{95} = \frac{N_{\bar{p}}}{\varepsilon_{L\,95} \cdot \varepsilon_{T\,n95}} \quad (1)$$

where $N_{\bar{p}}$ is the number of antiprotons in units of $10^{10}$, $\varepsilon_{L\,95}$ is the 95% longitudinal emittance in eV s and $\varepsilon_{T\,n95}$ is the normalized 95% transverse emittance in µm.

Then, for the Recycler, the instability threshold [5] can be written as follows:

$$D_{th,95} \approx 0.035 \cdot \xi_n F, \quad F \equiv \ln\left(\frac{\Delta v_{sc}}{\mathrm{Im}\,\Delta v_c}\right); \quad (2)$$

where $\xi_n = |\xi - n\eta|\sigma_p$ is the effective chromatic rms tune spread for mode $n$, $\eta$ is the slippage factor, $\sigma_p = \Delta p_{rms}/p$ is the relative rms momentum spread, $\Delta v_{sc}$ is the maximal space charge tune shift, and $\Delta v_c$ is the wake-driven coherent tune shift (see *e. g.*: [6]) believed to be dominated by the resistive wall contribution in RR. Since the effective chromaticity and the coherent tune shift depend on the mode frequency, or the harmonic number $n$, so does the instability threshold. For a 70 MHz damper system in the Recycler ($n \sim 780$), $F \sim 12$.

In the Recycler, the instability occurs primarily in the vertical direction while the model does not explicitly differentiate between the two transverse planes. However, the RR vertical resistive wall impedance is a factor of 2 higher than the horizontal, making the vertical threshold slightly lower due to the logarithmic factor $F$ for otherwise identical chromaticities and damper bandwidths.

The threshold expression (2) should be used with some caution. The coherent motion is stabilized by resonant particles, whose individual lattice tune shift compensates their individual space charge tune shift. For the space charge dominated impedance, these particles are in the far tails – longitudinal and transverse - of the beam distribution. When electron cooling is applied, there is no reason to assume the distribution to be Gaussian, and Eq. (2) is an approximation.

*Bunching effects*

Even if the synchrotron tune is much smaller than the coherent tune shift, there are at least three different ways for which beam bunching may influence the coherent oscillations.

First, the tail-to-head interaction is looped due to a long-range wake field that leads to a weak dependence of the coherent tune shift $\Delta v_c$ on the bunching factor $B = \tau_0/T_0 \leq 1$ (*i.e.* bunch length over revolution time). This dependence enters the expression of $\Delta v_c$ as $B^{-1/3}$ in Ref. [8] and $B^{-1/4}$ in Ref. [7]. Since $\Delta v_{sc} \propto 1/B$, Eq. (2) predicts a logarithmic growth of the stability threshold for shorter bunches.

The second effect from bunching appears when the RF barriers are lower than the maximum momentum offset of the resonant particles $\Delta p_{res\,max} = \left|\frac{\Delta v_{sc}(0)}{\xi_n} p\right|$. High-momentum antiprotons spend most of their time outside of the bucket and cannot effectively contribute to Landau damping. It makes the beam less stable than it would be

---

with a deeper potential well. This effect leads to a decrease of the instability threshold for shorter bunches.

A third factor is an increase of the beam stability threshold for RF configurations with smooth walls in comparison with the barrier configuration [9]. For example, a cosine-like potential well in which the beam is kept before extraction.

Finally, the presence of multiple bunches around the Recycler also logarithmically decreases the instability threshold.

## OBSERVATIONS

A total of six instabilities were observed after the final upgrade of the dampers (December 2007), all during extraction to the Tevatron. The extraction process includes complicated manipulations in the longitudinal phase space (*e.g.*: Figure 1). First, the bunch is divided into 9 nearly identical pieces by narrow rectangular barriers (called for historical reasons "mined bunches"). Then antiprotons are moved, one mined bunch at a time, into the extraction region. Once there, the mined bunch is adiabatically transformed into four 2.5 MHz smaller bunches, which are then extracted into the matching MI RF waveform. A detailed description of these manipulations can be found in [10].

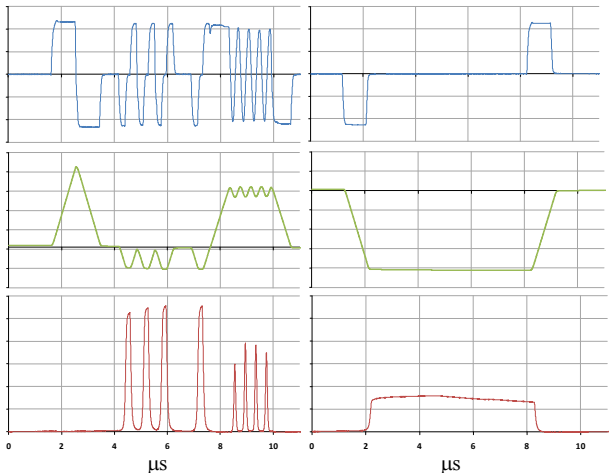

Figure 1: RF fan back (top row), equivalent potential (middle row) and beam linear density from the Resistive Wall Monitor (bottom row) taken in the middle of the extraction process (left) and the normal single bunch storing configuration (right). On the left plots, both the "mined bunches" and 2.5 MHz bunches are shown.

All of the instabilities were similar; a typical case is displayed in Figure 2, which shows the relevant parameters such as the antiproton intensity, transverse emittances and dampers kick amplitudes. The main characteristics are as follows:
- The beam loss occurs during the second half of the extraction process.
- Only one mined bunch at a time goes unstable.
- Typically, after the first instability, all remaining bunches become unstable as well at later stages. In a couple of exceptions, the very last bunch (#9) remained stable. In those cases, the bunch #9's intensity was ~20% lower than other bunch intensities because of imperfections of the RF system.
- Each beam loss lasts 5 – 15 seconds.
- The dampers act primarily in the vertical direction.
- There is a large emittance growth primarily in the vertical plane.

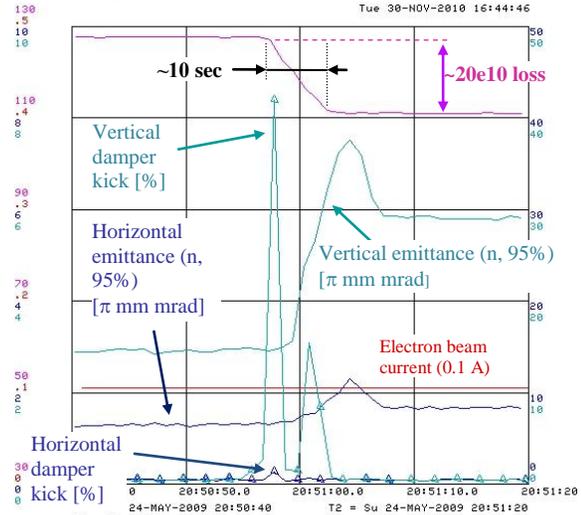

Figure 2: Instability during antiproton extraction to the Tevatron. The instability occurred on bunch #8, after extraction of bunch #6 (out of 9).

In addition, an oscilloscope was connected to the dampers pickup electrodes. It was triggered by a high transverse signal if it occurred above 70 MHz and recorded 32 ms of data. An example is shown on Figure 3 where traces were recorded during the same instability event as in Figure 2.

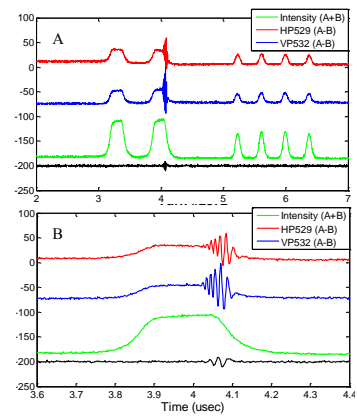

Figure 3: Oscilloscope traces from the damper pickups. The vertical scale is arbitrary. The green trace is the sum signal and is proportional to the linear density distribution. The red and blue traces are the difference signals (not normalized) for two damper pickups (red: horizontal; blue: vertical) and reflect the beam transverse position. The black trace is the damper vertical kick amplitude. Plots show the last turn of the recording interval (32 ms). A- ~1/2 of the RR circumference; B- zoom on the bunch going unstable.

In order to understand better the nature of these instabilities and determined experimentally the instability phase density threshold for the relevant RF configurations, various studies were carried out. In particular, one was designed in such a way that each measurement was mimicking one of the RF waveforms encountered during the extraction process.

## DATA SUMMARY AND DISCUSSION

The instability thresholds recorded during the studies are summarized in Table 1.

Table 1: Summary of the experimentally observed instability thresholds for the beam in a single bunch storing configuration (Figure 1).

|  | $D_{th,95}$ |
|---|---|
| No dampers ($n = 0$) | 0.5 – 0.8 |
| 30 MHz dampers ($n \sim 330$) | 2.6 - 3.1 |
| 70 MHz dampers ($n \sim 780$) | 3.4 – 6.2 |

While, the data have a large scatter and depart from the numbers calculated with Eq. (2) - 1.3 to 3.7 to 5.5, respectively - the model correctly describes the trend of introducing dampers with different bandwidths, and the observed thresholds are within a factor of two of the predicted numbers.

At nearly identical phase densities, studies show that the bunch structure the most prone to instabilities is the "mined bunch", in agreement with the operational observations. This peculiarity was explained by the combination of the high linear density and low barrier height (8.5 MeV/c vs standard 17 MeV/c) of this configuration. As it was mentioned previously, this leads to an effective exclusion from Landau damping of antiprotons with high longitudinal, low transverse actions. A dedicated study found that lowering the height of the barrier potential by a factor of two, mimicking what happens for the mined bunches, decreases the threshold phase density $D_{th,95}$ from 6.2 to 3.4.

Several features such as a slow non-exponential growth of the oscillations and seconds-long times beam losses were originally unexpected. However, a classical exponential growth of an instability describes the behaviour of a system sufficiently above the threshold, while in all our experiments the beam was slowly reaching the threshold density as it was being cooled. Strictly speaking, the instability growth rate at the exact threshold is zero. Then, in this case, it is determined not only by the impedance, but also by such factors as beam cooling, synchrotron motion and all sorts of diffusion for the resonant particles. That is why for that gradual approach of the threshold, the emerging instability can be orders of magnitude slower than the pure impedance-related growth.

## CONCLUSION

While the transverse dampers permitted to increase the maximum beam phase density by an order of magnitude, the transverse instability of the antiproton beam in the Recycler is the final limiting factor to the brightness of the extracted beams that can be achieved.

Qualitative features of the observed instances of the instability can be explained by the model developed for a coasting beam. The threshold phase is in agreement with the model within the scatter of experimental data and the precision to which this theoretical threshold can be calculated. The scatter in the data is likely related to variations in the distribution of the tails particles participating in Landau damping. In particular, lowering the potential depth of the barrier bucket effectively excludes part of the longitudinal tails from damping and may decrease the threshold phase density by a factor of two.

## ACKNOWLEDGMENTS

Authors acknowledge the participation of K. Seiya in several of the measurements.

## REFRENCES

[1] G. Jackson, "*Recycler Ring Conceptual Design Study*", FERMILAB-TM-1936 (1995)
[2] L.R. Prost *et al.*, Proc. COOL'09, Lanzhou, China (2009) MOM1MCIO01
[3] N. Eddy, J.L. Crisp, and M.Hu, AIP Conf. Proc. **868**: 293-297 (2006)
[4] N. Eddy, J. L. Crisp and M. Hu, Proc. EPAC'08, Genoa, Italy (2008) THPC117
[5] A. Burov & V. Lebedev, PRSTAB **12** 034201 (2009)
[6] A. Chao, "*Physics of collective beam instabilities in high energy accelerators*", J. Wiley & Sons, Inc, 1993
[7] A. Burov & V. Lebedev, AIP Conf. Proc. **773**: 350-354 (2005)
[8] V. Balbekov, PRSTAB **9** 064401 (2006)
[9] A. Burov, PRSTAB **12**, 044202 (2009)
[10] C.M. Bhat, Physics Letters A **330** (2004) 481-486